\documentclass[fleqn,10pt]{wlscirep}
\usepackage[utf8]{inputenc}
\usepackage[T1]{fontenc}
\usepackage{float}

\title{Beyond words and actions: Exploring Multimodal Analytics and Collaboration in the Digital Age}

\author[1,*]{Diego Miranda}
\author[1]{Rene Noel}
\author[1]{Jaime Godoy}
\author[1]{Carlos Escobedo}
\author[2]{Cristian Cechinel}
\author[1]{Roberto Munoz}
\affil[1]{Escuela de Ingenier\'ia Inform\'atica, Universidad de Valpara\'iso, Valpara\'iso, Chile.}
\affil[2]{Centro de Ci\^encias, Tecnologias e Sa\'ude, Universidade Federal de Santa Catarina, Araranguá, Brazil}

\affil[*]{diego.miranda@uv.cl}

\begin{abstract}

This article explores Multimodal Analytics' use in assessing communication within agile software development, particularly through planning poker, to understand collaborative behavior. Multimodal Analytics examines verbal, paraverbal, and non-verbal communication, crucial for effective collaboration in software engineering, which demands efficient communication, cooperation, and coordination. The study focuses on how planning poker influences speaking time and attention among team members by utilizing advanced audiovisual data analysis technologies. Results indicate that while planning poker doesn't significantly change total speaking or attention time, it leads to a more equitable speaking time distribution, highlighting its benefit in enhancing equitable team participation. These findings emphasize planning poker's role in improving software team collaboration and suggest multimodal analytics' potential to explore new aspects of team communication. This research contributes to better understanding coordination techniques' impact in software development and team education, proposing future investigations into optimizing team collaboration and performance through alternative coordination techniques and multimodal analysis across different collaborative settings.

\end{abstract}
\begin{document}

\flushbottom
\maketitle
%
%
\thispagestyle{empty}

\noindent Please note: Abbreviations should be introduced at the first mention in the main text – no abbreviations lists. Suggested structure of main text (not enforced) is provided below.

\section*{Introduction}

Multimodal Analytics (MmA) is an approach that enables the quantitative measurement of various elements of communication~\cite{Scharl2019, Riquelme2019, Koutsombogera2014}, covering verbal, paraverbal, and non-verbal aspects. This method provides data and visualizations that facilitate understanding the behavior of subjects evaluated in various contexts. By using MmA, it is possible to obtain a detailed view of communication components that otherwise might remain hidden or would require deep and laborious analysis of video recordings by experts. Some communicational aspects that have been addressed with MmA include the detection of body postures~\cite{Grover2016, Cukurova2018}, facial expressions~\cite{Muller2018, kim2023multi, Xu2023}, and verbal interventions~\cite{Riquelme2019, Lubold2014, Xu2023}, among others.

Communication is key to collaborative work, which is one of the cornerstones of creative multidisciplinary work. In particular, in the field of software engineering, agile methods~\cite{beck2001} have placed collaboration at the center of the software development process, considering it in the planning of a development sprint, in the daily evaluation of progress, in the review of results, and in the retrospective analysis of the performance of the development team~\cite{schwaber2002}. However, collaboration does not spontaneously occur simply by bringing people together around a goal. The conceptualizations of the domain~\cite{Souza2020, Castaer2020} have characterized collaboration as the result of effective \textit{communication}, \textit{cooperation}, and \textit{coordination}. \textit{cooperation} refers to the contribution of the different participants from their assigned roles and their knowledge and experience. However, even having experts with assigned roles, collaboration also depends on \textit{coordination}, that is, an agreement that defines how participants will interact in the collaborative activity. In agile methods, in particular, different coordination dynamics have been proposed, which, in theory, facilitate team collaboration. An example of these techniques is \textit{planning poker}~\cite{mahnivc2012using}, which defines a form of interaction for the participants of a development team to estimate the effort for the development of user stories, representing the features that the software should have. However, there is no scientific evidence on the effectiveness of these interaction dynamics, as their effect must be evaluated on a third component that, so far, has been elusive to measure quantitatively: \textit{ communication}.

Communication is related to other concepts and skills, such as collaboration and cooperation, which play a crucial role in building strong interpersonal relationships and enhancing the efficiency and effectiveness of teamwork~\cite{Sharp2010, Chassidim2018}. Additionally, these competencies are recognized as skills of the 21st century~\cite{vanLaar2017, Johnson_Johnson_2014, Fajaryati2020} and are highly sought after by employers~\cite{Askari2020, ElSofany2014}. However, effective communication is not an easy skill to achieve, as it does not occur merely when a group of people meets to talk to them. Instead, it requires that participating members are committed to the progress and achievement of the team, implying both cognitive and motivational dedication~\cite{Dillenbourg2009}. Studies have shown that within a classroom context, students do not know how to communicate effectively, especially at the beginning of these sessions~\cite{doi:10.3102/0034654318791584, doi:10.1080/0305764X.2016.1259389}, making it necessary to support students in improving their communication skills. The presence of a teacher or facilitator is required to guide learning and interactions, allowing students to articulate important ideas and participate in more meaningful exchanges~\cite{doi:10.3102/0034654318791584, Eshuis2019}.

It is noteworthy that non-verbal communication plays an integral role in these interactions, and it is crucial to acknowledge that it often gets overlooked, especially during teamwork. Nonverbal communication includes gestures, facial expressions, body language, and tone of voice, among others~\cite{Ridao2017}. These elements can convey a wealth of information and often complement or even contradict our words~\cite{GOSTAND1980, Asan2015}. However, in a team working environment, nonverbal communication can be difficult to perceive and easy to ignore~\cite{Ellgring1981}.

Considering how important and traditionally difficult it has been to objectively evaluate communication, MmA emerges as a possibility to conduct quantitative collaboration studies. According to the conceptualization of collaboration~\cite{Souza2020}, the application of MmA allows the combination of nonverbal and paraverbal elements to evidence changes in communication produced by different coordination agreements in teams while keeping the definitions of roles and responsibilities (cooperation) stable. The purpose of this study is to evaluate the effect of using a coordination dynamic on the time and attention of participants in a collaborative activity, measured with MmA techniques. The context in which the study is conducted involves agile software development teams that estimate the effort of a group of user stories, which must be achieved through a collaborative agreement. Each team first performs an effort estimation without a defined form of coordination and then, using planning poker, in two different but similar problems, maintaining their participants' roles. During both sessions, audio and video are recorded and analyzed with MmA techniques to measure speaking time and attention while speaking. Speaking time is measured using diarization techniques to discriminate the oral interventions made by each participant, and attention is measured using the orientation of the face among participants during such interventions. After the experimental sessions, charts and visualizations consolidating the MmA measurements were presented to domain experts, who generated hypotheses from their observations.

\section*{Methods}

Our research aims to study the effect of a coordination dynamic on collaboration, leveraging the potential of MmA to measure nonverbal and paraverbal aspects of communication. It is situated in the specific context of collaborative activities for agile software development~\cite{Cockburn2001, Alsaqqa2020}, particularly the use of planning poker~\cite{grenning2002planning} for sprint planning. We define the study according to the guidelines presented by Wohlin et al.~\cite{Wohlin2014}.

Analyzing the planning poker technique with the purpose of evaluating its effect on non-verbal communication from the researcher's perspective in the context of undergraduate students performing a co-located collaborative activity.

The research questions for the study are two, detailed below along with the respective null hypotheses.

\begin{itemize}

\item RQ1: What is the effect of using poker planning on the speaking time of the participants in the collaborative activity? The associated null hypothesis is $H_0st$: Planning poker has no effect on the speaking time of participants.

\item RQ2: What is the effect of using poker planning on the attention time of the participants in the collaborative activity? The associated null hypothesis is $H_0at$: Planning poker has no effect on the attention time of participants.

\end{itemize}

For the first two research questions, which require measuring speaking time and attention time, the application of the MmA approach is fundamental. For this reason, the following subsection will dive into how data are collected and processed to operationalize these variables, then continue with the design of the experimental activity and data analysis.

\subsection*{Data collection} 

For this study, data collection was carried out using a multimodal methodology, focused on capturing participant interactions through audio and video recordings. Advanced technologies were utilized to accurately capture the dynamics and interactions within the teams. Sessions were recorded using a Kandao Meeting Pro camera system, which includes a high-definition camera with the capability of recording in 360-degree panorama, ensuring a complete view at the center of the activity.

\subsubsection*{Audio data} 

The audio analysis focused on understanding and quantifying the verbal interaction among participants, a critical component for assessing communication dynamics in collaborative environments~\cite{Praharaj_2021}. By utilizing advanced voice recognition technologies, we aimed to extract accurate data on speaking time and participant diarization.

The central tool in this process was WhisperX~\cite{Bain2023}, an advanced, open-source version of the renowned Whisper voice recognition system~\cite{10.5555/3618408.3619590}. WhisperX is distinguished by its temporal precision, offering word-level segmentation through vocadetection of vocal activity forced phoneme alignment. This capability significantly enhances the identification and allocation of audio segments to the corresponding speakers, a process known as diarization. Diarization is crucial for determining who is speaking and when, allowing for a detailed analysis of the group's communication dynamics.

To ensure the quality and accuracy of the data, all audio files were converted to WAV format and processed using the Large-v3 model, which presents a lower word error rate. The resulting transcriptions were stored in JSON format, facilitating their analysis and integration with visual data.

Manual review of each file was a critical step in ensuring the accuracy of the information. This meticulous process involved verifying timestamps, correcting speaker misidentification, and removing incorrectly assigned audio segments. This rigorous approach ensured that the analyzed data reliably reflected verbal interactions during the sessions, minimizing distortions and providing a solid foundation for further analysis.

\subsubsection*{Video data}

\paragraph{Participant Detection and Tracking:}

\begin{figure}[ht!]
    \centering \includegraphics[width=\textwidth,height=\textheight,keepaspectratio]{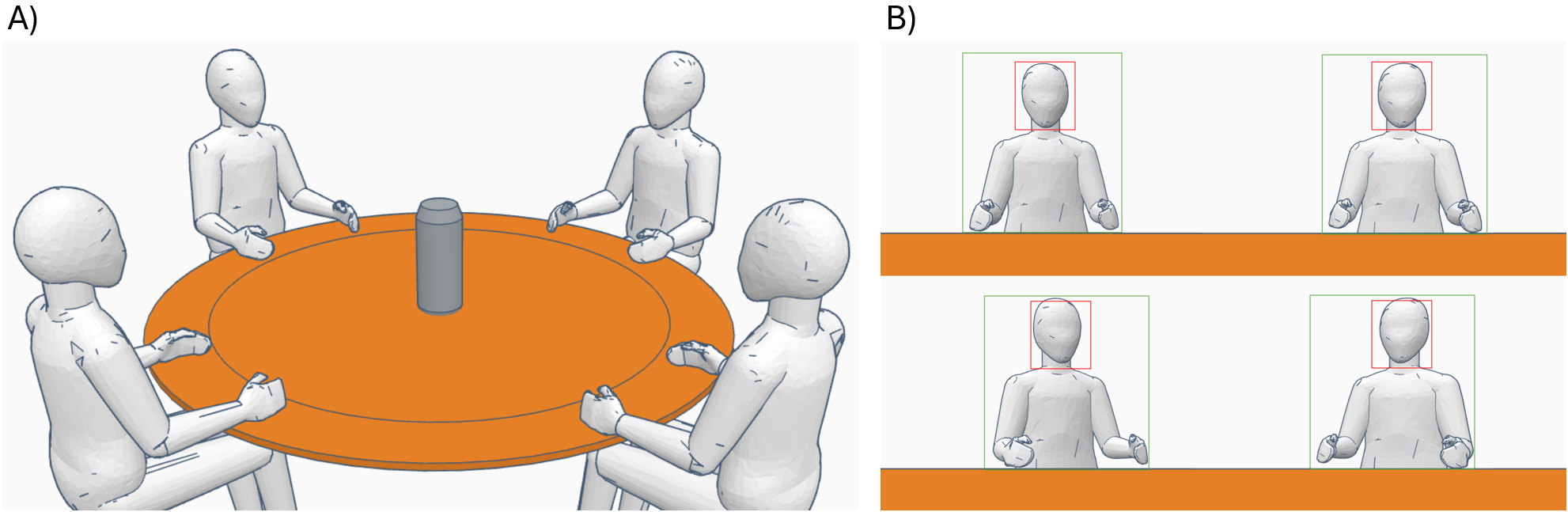}
    \caption{A) Example of a group activity viewed from the outside. B) Example of a group activity viewed with the 360-degree camera. The bounding box of the participant is indicated in green. The bounding box for the participant's face is indicated in red.}
    \label{fig:Experiment}
\end{figure}

Initially, we applied YOLOv8~\cite{Jocher_Ultralytics_YOLO_2023} to accurately identify each participant in the video recordings. This real-time object detection model excels in recognizing and labeling individuals quickly, generating bounding boxes around each one, marked in green in the visualization of Figure~\ref{fig:Experiment}.B. Subsequently, we used DeepSORT~\cite{Wojke2018deep}, through the Python library \textit{deep\_sort\_realtime}, to track the movement of the participants throughout the session. This combination of detection and tracking ensures the precise correlation of individuals with their specific actions.

\paragraph{Facial Landmark Detection:}

\begin{figure}[ht!]
    \centering \includegraphics[width=\textwidth,height=\textheight,keepaspectratio]{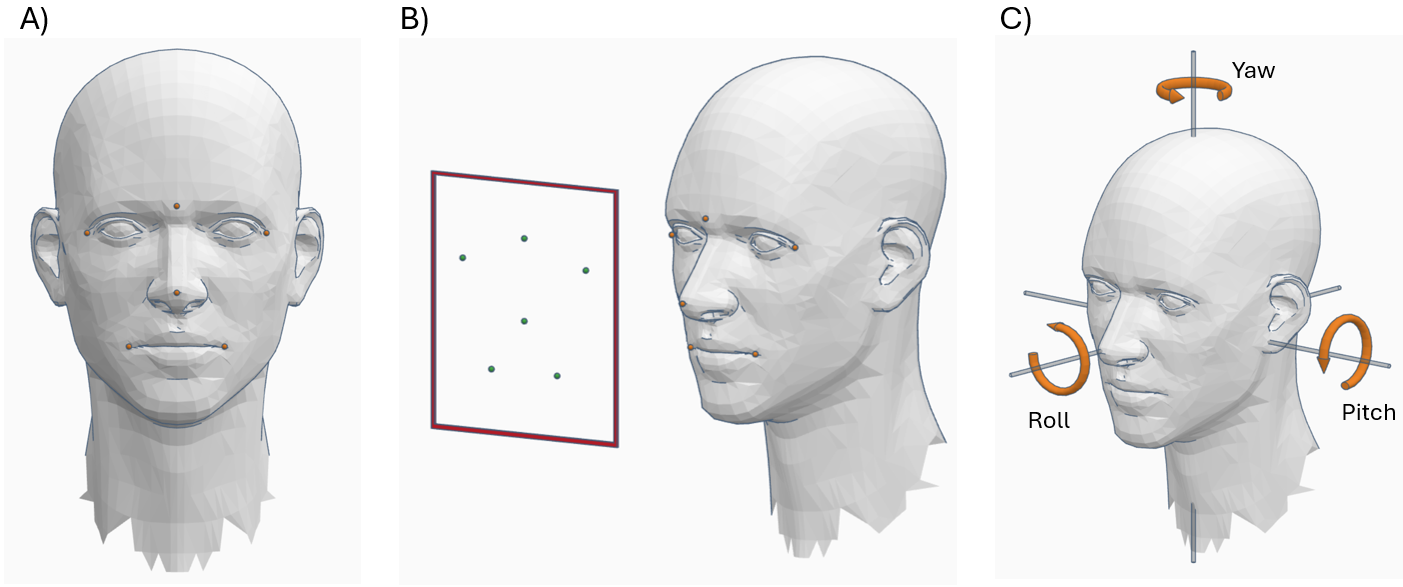}
    \caption{A) Example of the facial points used to estimate its direction. B) Example of 2D points relative to the 3D points of the face. C) Diagram of the rotation axes $pitch$, $yaw$ y $roll$. }
    \label{fig:Estimation}
\end{figure}

With the participants clearly identified, we used facial landmarks detected by MediaPipe~\cite{mediapipe} to detect the face and capture the fine details of non-verbal communication (bounding box in red, from Figure~\ref{fig:Experiment}.B). The landmarks are 1, 9, 57, 130, 287, and 359 according to the MediaPipe documentation, illustrated in Figure~\ref{fig:Estimation}.A. These points are crucial for assessing the orientation of the face and, therefore, the attention, interest, and participation of subjects during group interaction.

\paragraph{Accurate Face Direction Estimation:}

The direction of the face is estimated using the solvePnP (Perspective-n-Point) function of OpenCV~\cite{opencv_solvepnp}, which calculates the three-dimensional orientation of the face based on known facial landmarks. This process involves transforming these points from a 3D model to their corresponding location in the 2D image (see Figure~\ref{fig:Estimation}.B). From this function, we obtain a rotation vector that is converted into a rotation matrix to determine the face orientation axes: $pitch$, $yaw$, and $roll$ (see Figure~\ref{fig:Estimation}.C).

By obtaining the face orientation, having a camera with panoramic vision, and all participants at the same distance, we can estimate towards which other participant their face is pointing. This is done considering that the camera has 45 and 360 (2 images of 180) degrees of vision on the vertical and horizontal axes, respectively, and the participants are at the same distance from the camera. We convert the point of the nose (1) into a vector $\vec{v}$, whose angle represents the location of the participants' faces with respect to the camera's coordinate axis.

$$\vec{v} = r(cos(a)x + sin(a)y + tan(b)z)$$

$a$ y $b$  are the vertical and horizontal angles of the nose, respectively, and $r$ is the distance of the participants from the camera.

Having obtained the vector of the location of each participant and the direction of the face, the focus vector $\vec{v_f}$ that indicates where a participant is directing their face, from the camera's perspective, is calculated. This calculation is based on the following formulas:

$$lx = r + r \cdot \cos(2 \cdot yaw)$$
$$ly = r \cdot \sin(2 \cdot yaw)$$
$$h = \sqrt{lx^2 + ly^2}$$
$$lz = h \cdot \tan(b - pitch)$$

$$
\vec{v_f} = \begin{bmatrix}
-\vec{v}_x \cdot \cos(2 \cdot yaw) - (-\vec{v}_y \cdot \sin(2 \cdot yaw)) \\
-\vec{v}_x \cdot \sin(2 \cdot yaw) + (-\vec{v}_y \cdot \cos(2 \cdot yaw)) \\
-\vec{v}_z + lz\\
\end{bmatrix}
$$

\paragraph{Detailed Analysis of Attention:}

Finally, having obtained the focus vector of all participants, the reverse process is conducted to obtain the angle with respect to the camera. At this point, to determine whom a participant is observing, 2 thresholds were established. The first threshold (vertical axis) corresponds to whether the person is holding their head up or looking at documents. This threshold was determined to be 15 degrees (out of a total of 45 degrees), which means that if the vertical angle is below 15 degrees, the participant is reading; otherwise, they could be observing another participant. The second threshold (horizontal axis) was determined to be 25\%, which means that if participant $A$ has participants $B$, $C$, and $D$ in front, to the right, and to the left, respectively, the degree distance regarding the camera from participant  $B$ to  $C$ and $D$ will be calculated, then
25\% of this distance is calculated.

$$u_r = (|angle(C)| - |angle(B)|) * 0.25$$
$$u_l = (|angle(B)| - |angle(D)|) * 0.25$$

Let $u_r$ be the threshold to the right of participant $B$ and $u_l$ the threshold to the left of the participant. Then, if the horizontal angle of the focus vector of $A$ is less than the threshold $u_r$, it is determined that $A$ is paying attention to $C$. If the horizontal angle is greater than the threshold $u_l$, it is determined that $A$ is paying attention to $D$. When the horizontal angle is between the threshold $u_r$ and $u_l$, it is determined that $A$ is paying attention to $B$.

With the aforementioned analysis, it is possible to determine if, in a specific frame of the video, a participant is paying attention to another, their faces falling within the previously described thresholds. By summing up the frames, it is possible to calculate the amount of attention time each participant has given. To determine whether attention is being paid during a verbal intervention, only the frames whose timestamp falls within the range of verbal interventions determined for the participant receiving the attention are considered. The times of verbal interventions are determined by diarization, as explained earlier in the audio detection subsection.

The final results are stored in CSV format, documenting the number of frames in which a participant directs their attention towards another.

\subsection*{Experimental Design}

\subsubsection*{Factors and Variables}

The factor under study is the coordination technique, which has two levels: ad hoc, or without prior coordination instructions, and planning poker. The dependent variables associated with the three research questions and the metrics that operationalize them are described below.

\begin{itemize}
    \item \textbf{Speaking Time:} Refers to the time that each participant speaks during the activity. Three metrics were calculated: total speaking time for all participants, average speaking time among participants, and standard deviation of speaking time for each participant.
    
    \item \textbf{Attention Time:} Refers to the attention time of each participant during verbal interactions. The direction of attention was analyzed based on the orientation of the face, determining toward which other participants this attention was directed and at what moments. With these data, three metrics were calculated: the total attention time of all participants, the average attention time of each participant, and the standard deviation of attention time for each participant. 
\end{itemize}

The subject of the study is a collaborative group composed of four participants. Each group participates in two experimental activities; in the first, they collaborate without a coordination technique (Activity A), and in the second, they use planning poker (Activity B), providing a uniform comparative basis. Teams have a maximum of 10 minutes to reach an agreement. The agreement consists of rating the complexity of developing a user story that describes a software functionality in a 2-week sprint. User stories, defined by their complexity level, are classified using a five-level Likert scale. This scale ranges from \textit{Very Simple}, where the problem requires minimal effort and can be quickly completed within the Sprint, to \textit{Impossible}, indicating that the user story exceeds the team's capacity to be developed within the assigned Sprint time. The intermediate levels, \textit{Simple}, \textit{Medium}, and \textit{Complicated}, offer varying degrees of difficulty and possibilities for including additional work in the Sprint, thus providing a realistic spectrum of challenges that software development teams regularly face.

The experiment considers keeping certain independent variables constant throughout the experiment to block their effect. In particular, the composition of the team, the type and complexity of the estimated user stories, and the physical and technological environment in which the sessions are conducted, to ensure the validity and reliability of the results. Additionally, each participant has a unique role within the team from which they must cooperate, which can be Backend, Frontend, UI/UX, and data persistence. This allows isolating the effect of introducing planning poker on the interest variables.

During the experiments, participants were positioned around a circular table, facilitating precise detection and recording of interventions and behaviors, such as visual focus and attention to other members. Each participant received a detailed introduction to the problem to be solved and a diagram of the related architecture, ensuring equality of information in the discussions.

\subsection*{Participant selection.}

During the months of August and November 2023, a public call was extended to students of the School of Computer Engineering at the University of Valparaíso to voluntarily participate in recording sessions. Students interested in participating were required to have previously completed the course on Fundamentals of Software Engineering. This requirement was established to ensure that all participants had a common basic knowledge related to software development and agile methodologies.

We received approval from the Institutional Bioethics Committee for Research on Human Subjects of the Universidad de Valpara\'iso (protocol code CEC-UV 236-21) and conducted according to the guidelines of the Declaration of Helsinki. Participants gave informed consent before joining the study, ensuring their understanding and agreement with the terms of their participation. No financial compensation was offered for their participation, highlighting the voluntary nature and personal interest in contributing to the advancement of research in software engineering and team dynamics.

Group formation was based on the participants' preferences, who had the freedom to choose whether they wanted to collaborate with people they already knew or with new peers. This flexibility allowed for diversity in the composition of the teams, reflecting real-life team formation situations in educational and professional environments. Each group consisted of four students, achieving a total participation of 72 students, which resulted in the formation of 18 different teams.

To document and analyze the development of the activities and the dynamics of the teams, detailed information on the gender distribution of the participants, the time dedicated to each activity, and the level of agreement reached by each team was collected. This information is summarized in Table 1 (Table~\ref{tab:activity-info}), which provides an overview of the team dynamics and the outcomes of the activities under the different experimental conditions.

\subsection*{Data Analysis.}
To answer research questions, an analysis was performed using statistical and computational tools to process and analyze the data sets obtained from audio and video recordings, allowing an objective and quantitative interpretation of communication and collaboration dynamics within teams. Statistical analysis was carried out following these steps:

\begin{itemize}
    \item Data Inspection: Box plots are created for each measurement to study the presence of outliers in the collected data, as these extreme values can significantly influence the underlying assumptions of many statistical tests. To quantitatively identify outliers, the interquartile range (IQR) method is used. This method defines outliers as points below $Q_1 - 1.5$ or above $Q_3 - 1.5 \times IQR$, where $Q_1$ $ Q_3$ are the first and third quartiles, respectively, and $IQR$ is the interquartile range $(Q_3 -  Q_1)$.
    
    \item Verification of prerequisites for t-test: A comparison between the conditions with and without predefined coordination was performed to evaluate the impact of planning poker on the communication and collaboration of the teams. The Shapiro-Wilk test~\cite{SHAPIRO1965} was used to verify the normality of the data distribution for each measurement, and the Levene test~\cite{Levene1960} was used to verify the homogeneity of variances.
    
    \item Application of t-test: After verifying the necessary prerequisites, the t-test is applied to compare the difference between the measurements with and without coordination. Its statistical significance is verified if the p-value is less than or equal to $.05$.

    \item Calculation of Effect Size (Cohen's d): To complement the analysis with the t-test and gain a deeper understanding of the actual impact of the intervention, we calculate the effect size using Cohen's d. This indicator allows us to quantify the magnitude of the difference between the coordination and no-coordination groups, providing a measure of the practical relevance of the results. Cohen's d is calculated as the difference between the means of the two groups divided by the combined standard deviation of the groups. A d value of 0.2 is considered a small effect, $0.5$ a medium effect, and 0.8 or more, a large effect. This calculation helps us interpret the statistically significant differences observed and have practical relevance in the study context.
\end{itemize}

\subsection*{Threats to Validity Analysis}

Regarding threats to the validity of this study, it is important to mention several aspects that could influence the interpretation and generalization of the results. One threat is the learning effect, which may arise when participants are exposed to both experimental conditions (with and without planning poker). This threat is associated with the risk that the experience gained in the first session influences performance in the second, regardless of the coordination technique used. To mitigate this threat, the experiment's design included changing the user stories between sessions and alternating the conditions for each group. However, participants were not informed of the specific aspects being measured, which reduces the possibility that they intentionally modified their behavior to adapt to the perceived expectations of the study.

Another potentially relevant threat is participant selection. Since participation in the study was voluntary and targeted at a specific group of computer engineering students with a certain knowledge of agile methodologies, the results may only be generalizable to some contexts of agile software development or teams with different experience levels. Although the nonrandom selection could limit the generalization of the findings, the diversity in team formation, based on participants' preference to collaborate with known or new peers, attempts to reflect varied team dynamics found in educational and professional environments.

Finally, another domain of threats to validity concerns the definition and measurement of variables. Although this study sought to apply a rigorous and multimodal approach to capture communication and attention dynamics in teams, caution is required in interpreting the collected data. It is possible that the metrics of speaking time and attention do not fully capture the complexity of these interactions in a team-working environment. Future research could incorporate additional measures, such as analysis of communication content or evaluation of the quality of work produced, to better understand how coordination techniques influence effective collaboration.

\section*{Results}
A total of 18 groups and 72 participants completed the experimental activities. The results of the gender distribution of the participants in each group, the duration of the activity and the outcome of the complexity estimation are shown in Table~\ref{tab:activity-info}.

\begin{table}[ht!]
\centering
\resizebox{\columnwidth}{!}{%
\begin{tabular}{l|ll|ll|ll|}
\cline{2-7}
                            & \multicolumn{2}{l|}{Number of participants} & \multicolumn{2}{l|}{Time spent (min)}    & \multicolumn{2}{l|}{Level of complexity} \\ \hline
\multicolumn{1}{|l|}{Group} & Male                & Female                & Without coordination & With coordination & Without coordination   & With coordination   \\ \hline
\multicolumn{1}{|l|}{1}  & 3 & 1 & 8:10  & 9:30  & Medium        & Complicated   \\
\multicolumn{1}{|l|}{2}  & 4 & 0 & 10:00 & 9:30  & Medium        & Complicated   \\
\multicolumn{1}{|l|}{3}  & 4 & 0 & 7:34  & 10:00 & Simple        & Simple     \\
\multicolumn{1}{|l|}{4}  & 4 & 0 & 7:30  & 10:00 & Medium        & Complicated   \\
\multicolumn{1}{|l|}{5*}  & 4 & 0 & 9:46  & 9:45 & Medium        & Medium        \\
\multicolumn{1}{|l|}{6}  & 2 & 2 & 7:20  & 7:15  & Simple        & Simple     \\
\multicolumn{1}{|l|}{7}  & 1 & 3 & 3:40  & 4:27  & Simple        & Complicated   \\
\multicolumn{1}{|l|}{8}  & 4 & 0 & 7:47  & 3:30  & Impossible    & Simple     \\
\multicolumn{1}{|l|}{9}  & 4 & 0 & 10:00 & 5:00  & No agreement  & Simple     \\
\multicolumn{1}{|l|}{10**} & 4 & 0 & 9:35  & 9:10  & Complicated & Medium        \\
\multicolumn{1}{|l|}{11} & 4 & 0 & 9:32  & 9:49  & Complicated   & Complicated   \\
\multicolumn{1}{|l|}{12} & 2 & 2 & 10:00 & 6:15  & Very Simple   & No agreement \\
\multicolumn{1}{|l|}{13*} & 3 & 1 & 4:46 & 8:11  & Medium        & Impossible    \\
\multicolumn{1}{|l|}{14} & 3 & 1 & 9:10  & 6:00  & Complicated   & Complicated   \\
\multicolumn{1}{|l|}{15} & 4 & 0 & 10:00 & 7:09  & Medium        & Very Simple \\
\multicolumn{1}{|l|}{16} & 3 & 1 & 9:40  & 9:52  & Medium        & Medium        \\
\multicolumn{1}{|l|}{17} & 3 & 1 & 4:50  & 6:00  & Medium        & Simple     \\
\multicolumn{1}{|l|}{18} & 4 & 0 & 5:38  & 7:55  & Simple        & Medium        \\ \hline
\end{tabular}
}
\caption{Information for each group after the activity is completed. Note: An * indicates that participants were removed from the sample for not following the instructions of the experiment. A ** indicates that the data are considered outliers.}
\label{tab:activity-info}
\end{table}

After data collection, Groups 5 and 13 were discarded because it was verified during the activity that the participants did not follow the established procedure.

Below are the detailed results for each experimental variable.

\subsection*{Speaking Time}
\begin{figure}[ht!]
    \centering \includegraphics[width=0.85\textwidth,height=\textheight,keepaspectratio]{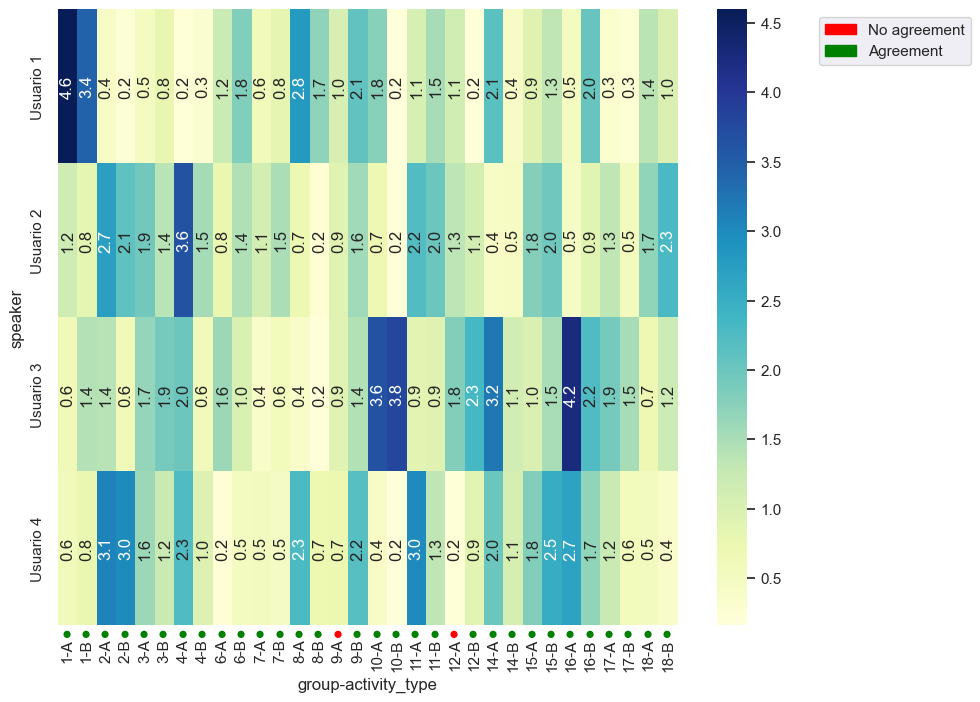}
    \caption{Heatmap of total speaking time in minutes per group and activity.}
    \label{fig:speakingtime_heatmap}
\end{figure}

The application of MmA allowed for the rapid generation of visualizations to guide the analysis of the produced data. In Figure~\ref{fig:speakingtime_heatmap}, a heatmap is shown comparing, for each group, the speaking time in their session without coordination (Activity A) and with coordination (Activity B). Through simple inspection, it is possible to hypothesize that the "heat" of the groups tends to moderate in Activity B compared to Activity A.

\begin{figure}[ht!]
    \centering \includegraphics[width=0.8\textwidth,height=\textheight,keepaspectratio]{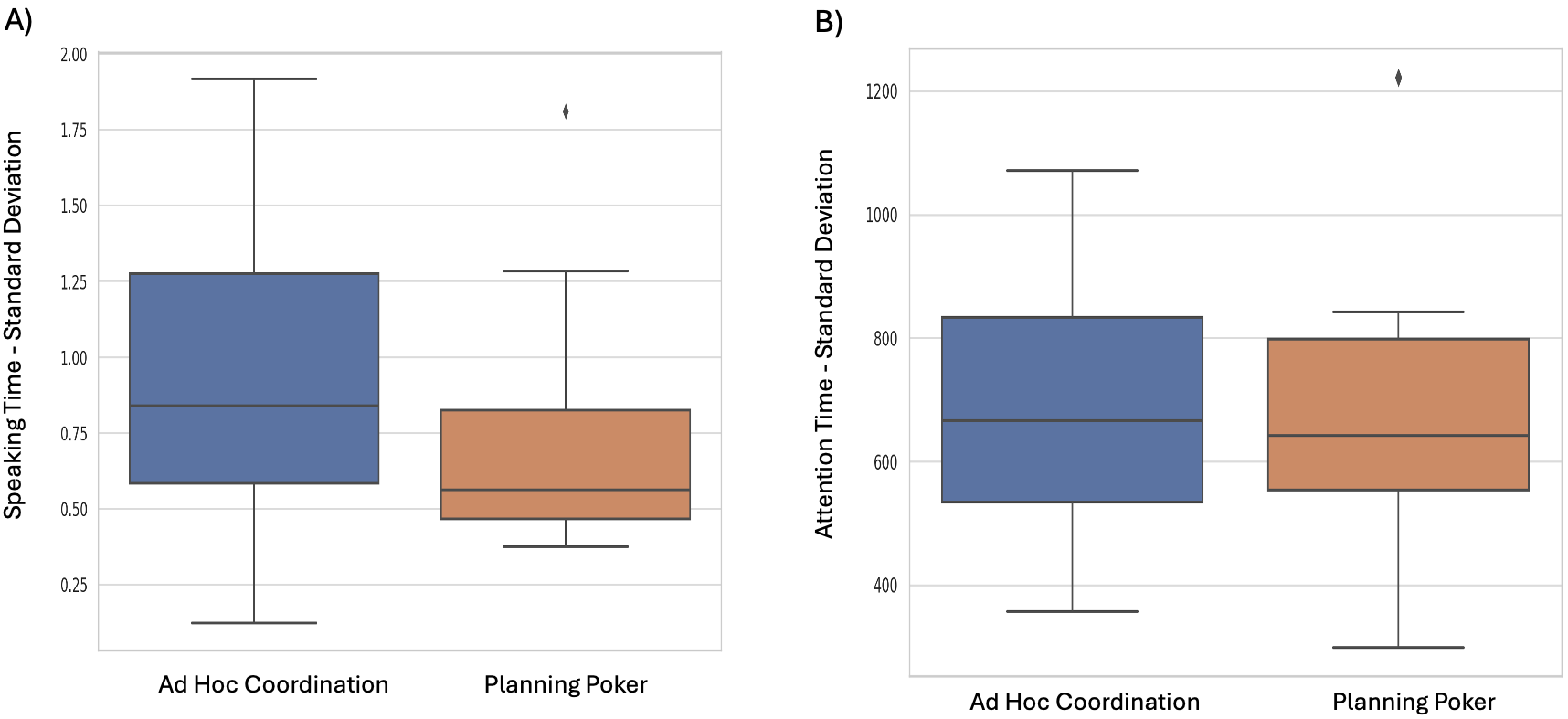}
    \caption{Boxplot for the Speaking Time Standard Deviation (STSD) Metric}
    \label{fig:boxplot}
\end{figure}

Outliers for speaking time across the three metrics: Total Speaking Time (TST), Average Speaking Time (AST), and Speaking Time Standard Deviation (STSD), for activities without coordination and with coordination, were reviewed. As seen in Figure~\ref{fig:boxplot}.A), the boxplot for STSD shows an outlier. After conducting the interquartile range analysis to identify the outlier and verifying in the video recording that group 10 had not followed the procedure, it was removed from the analysis.

The normality and homogeneity of variance tests were passed for all measurements, which allowed the application of the t-test to evaluate significant differences. The statistics of the tests are detailed in Table~\ref{tab:speakingtime}.

\begin{table}[ht]
\begin{center}
\caption{Speaking time statistics.}
\label{tab:speakingtime}
\begin{tabular}{lllll}
Variable & p-value (Test t) & Shapiro-Wilk   A & Shapiro-Wilk   B & Levene's test \\ \hline
TPH      & 0.139930           & 0.378962         & 0.237367         & 0.560153           \\
TTH      & 0.139930           & 0.378962         & 0.237367         & 0.560153           \\ \hline
DEH      & 0.050868           & 0.495974         & 0.002419         & 0.059034           \\

\end{tabular}
\end{center}
\end{table}
The results show that while the p-value of the t-test does not reach .05 for any of the measurements, for the Speaking Time Standard Deviation (STSD), its value is very close to statistical significance ($p = 0.560153$). The effect size of this difference is large (Cohen's $D = 0.610$), suggesting that the differences can be perceived in practice.

The interpretation of this result is that the coordination technique allows subjects who spoke less without defined coordination to now speak more, and those who spoke more, now speak less. This would democratize the use of speaking time during collaborative activities.

With this, the null hypothesis  $H_0st$ for the standard deviation of the participants' speaking time is rejected.

\subsection*{Attention Time}
\begin{figure}[ht!]
    \centering \includegraphics[width=0.8\textwidth,height=\textheight,keepaspectratio]{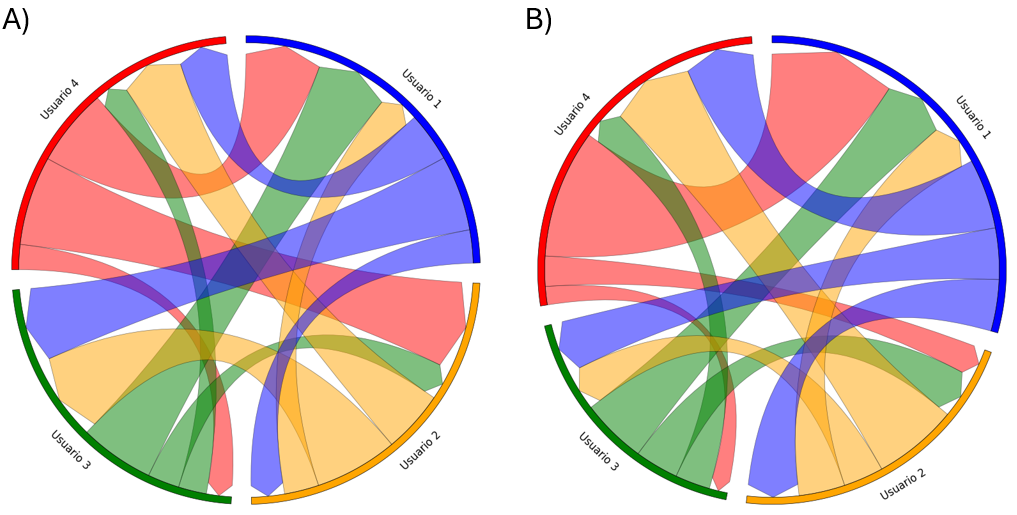}
    \caption{Attention time chart for Group 11, Activity A and Activity B. The length of the arc segment represents the speaking time of each participant, the width of the arrows represents the attention time during speaking, and the direction of the arrow indicates the user's attention to others (arrows from the participant) and the time the participant receives attention from others (arrows towards the participant).}
    \label{attention_time_chart}
\end{figure}

To explore the results of attention times, visualizations were developed for each group, like the one shown in Figure~\ref{attention_time_chart}, which account for the amount of time each participant pays attention to others while they are speaking, as well as the attention they receive while the subject speaks. The visualizations were compared with the video recording of each activity, finding minor errors in the automatic detection in the identification of users, which were corrected before the analysis of results. At first glance, no distinct attention patterns between the two activities each group participated in were identified, such as subjects who initially did not receive attention and then did, for example.

Outliers for the attention time across the three metrics: Total Attention Time (TAT), Average Attention Time (AAT), and Attention Time Standard Deviation (ATSD), for activities without coordination and with coordination, were reviewed. As seen in Figure~\ref{fig:boxplot}.B), the boxplot for ATSD shows an outlier. After conducting the interquartile range analysis to identify the outlier, it corresponds to Group 10, which was identified and excluded in the previous subsection.

\begin{table}[ht]
\begin{center}
\caption{Attention time statistics.}
\label{tab:Attentiontime}
\begin{tabular}{lllll}
Variable & p-value (Test t) & Shapiro-Wilk   A & Shapiro-Wilk   B & Levene's test \\ \hline
TPA      & 0.848904           & 0.980390         & 0.258090         & 0.953797           \\
TTA      & 0.848904           & 0.980390         & 0.258089         & 0.953797           \\ \hline
DEA      & 0.346346           & 0.272863         & 0.562062         & 0.329771           \\

\end{tabular}
\end{center}
\end{table}
Normality and homogeneity of variance tests were passed for all measurements, allowing the application of the t test to evaluate significant differences. The statistics for the tests are detailed in Table~\ref{tab:Attentiontime}.

The results of the Attention Time statistical tests indicate that no statistically significant differences were found between conditions with and without the use of the coordination technique for any of the metrics related to Attention Time. The p-values for the t-tests of these metrics exceed the statistical significance threshold of 0.05, implying that there is not enough evidence to reject the null hypotheses associated with these measures.

Specifically, for the ATSD, even after a detailed inspection of outliers and the exclusion of Group 10, the resulting p-value of the t-test ($0.346346$) remains far from the threshold of 0.05, indicating that there is no significant difference in the variability of attention time between the two coordination conditions. This suggests that the use of planning poker did not significantly affect the way participants distributed their attention during collaborative activities.

\section*{Discussion}

In relation to the first research question (RQ1) which explores the impact of employing the technique known as "planning poker" on the participation time of those involved, it is observed that, despite there being no statistically significant differences regarding the total intervention time, the standard deviation of speaking time exhibits a considerable decrease when this coordination methodology is implemented. This finding suggests that the use of the "planning poker" technique encourages a more equitable distribution of speaking time among participants, thereby contributing to greater equality in the allocation of intervention time during collaborative activities. Consequently, the null hypothesis $H_0st$ associated with the standard deviation of speaking time is rejected, confirming that "planning poker" significantly affects this variable.

Regarding the second research question (RQ2), which examines the impact of using "planning poker" on participants' attention time, the results reveal that no statistically significant differences were recorded in any of the metrics associated with attention time. This finding suggests that, although the "planning poker" methodology may influence the distribution of participation time by those involved, it does not significantly affect the way they pay attention during activities. Therefore, the null hypothesis
$H_0at$ is accepted, indicating that planning poker does not have a significant effect on the attention time of the participants.

The difference in results between speaking time and attention time could be explained by the specific nature of each type of interaction. While planning poker seems to facilitate a more equitable distribution of speaking time, possibly due to its structure that grants explicit turns for participation, the attention participants pay does not seem to be directly influenced by this coordination mechanism. This could be because attention is more susceptible to individual factors and group dynamics that are not simply altered by changing the way discussion is organized, such as participants' previous experience working together, their level of interaction, and team size, as reported by studies on team participation~\cite{kiani2013measuring}.

These findings are consistent with previous work that has explored how different collaboration techniques affect team dynamics in agile environments, suggesting that while some practices may improve specific aspects of collaboration, their impact can vary depending on the team characteristics and context. In~\cite{Haugen2006}, an empirical study is presented on the impact of using planning poker for user story estimation by a team using the Extreme Programming (XP) methodology. The comparison between an unstructured estimation process and the semi-structured use of "planning poker" revealed overall improvements in team performance, although an increase in estimation error was observed in extreme cases. The same case for~\cite{Poenel2023}, whose goal was to evaluate the effectiveness of effort estimation methods in software development. University students volunteered, who were not informed in detail about the study's objectives to prevent them from manipulating the results. It was found that some agile methods, one of them "planning poker," are more accurate and efficient in estimating the effort to complete user stories in a project.

Regarding effort estimation in agile methods, this work pays attention to factors not previously measured: evidence of the effects of estimation methods is usually measured based on the accuracy of the estimates~\cite{schweighofer2016effort}, however, the effective and equitable participation of team members had been unexplored until now. In this sense, the implications of this MmA application study to collaborative work transcend the agile software development context. The ability to identify differences in the spread of speaking time between participants could provide a powerful nonverbal communication indicator to assess, in real-time, situations of inequity in collaboration, which in educational contexts has been identified as a foundational component of social justice~\cite{patterson2019equity}.

In this research, most of the potential threats have been successfully mitigated through meticulous study design. Through the application of MmA techniques, we have been able to obtain detailed insights into the verbal, paraverbal and nonverbal aspects of communication in agile software development teams. These analyses have allowed us not only to evaluate the effectiveness of coordination techniques such as "planning poker" in the equitable distribution of speaking time among participants, but also to better understand how participants direct their attention during collaborative activities. Despite these advances, we recognize that the depth and reliability of our measurements can be further improved. Therefore, future work will aim to complement the multimodal analysis presented in this article with the detection of other aspects of non-verbal and verbal language, such as body posture identification and content analysis of oral interventions. The purpose of these measurements will be to contribute to the measurement and evaluation of the equitable participation of members in a collaborative work team.

\section*{Data availability}

The datasets generated and/or analyzed during the current study are available from the corresponding author upon reasonable request.
\bibliography{sample}

\begin{thebibliography}{10}
\urlstyle{rm}
\expandafter\ifx\csname url\endcsname\relax
  \def\url#1{\texttt{#1}}\fi
\expandafter\ifx\csname urlprefix\endcsname\relax\def\urlprefix{URL }\fi
\expandafter\ifx\csname doiprefix\endcsname\relax\def\doiprefix{DOI: }\fi
\providecommand{\bibinfo}[2]{#2}
\providecommand{\eprint}[2][]{\url{#2}}

\bibitem{Scharl2019}
\bibinfo{author}{Scharl, A.} \emph{et~al.}
\newblock \emph{\bibinfo{title}{Multimodal Analytics Dashboard for Story Detection and Visualization}}, \bibinfo{pages}{281–299} (\bibinfo{publisher}{Springer International Publishing}, \bibinfo{year}{2019}).

\bibitem{Riquelme2019}
\bibinfo{author}{Riquelme, F.} \emph{et~al.}
\newblock \bibinfo{journal}{\bibinfo{title}{Using multimodal learning analytics to study collaboration on discussion groups}}.
\newblock {\emph{\JournalTitle{Universal Access in the Information Society}}} \textbf{\bibinfo{volume}{18}}, \bibinfo{pages}{633--643}, \doiprefix\url{10.1007/s10209-019-00683-w} (\bibinfo{year}{2019}).

\bibitem{Koutsombogera2014}
\bibinfo{author}{Koutsombogera, M.} \& \bibinfo{author}{Papageorgiou, H.}
\newblock \bibinfo{title}{Multimodal analytics and its data ecosystem}.
\newblock In \emph{\bibinfo{booktitle}{Proceedings of the 2014 Workshop on Roadmapping the Future of Multimodal Interaction Research including Business Opportunities and Challenges}}, ICMI ’14, \doiprefix\url{10.1145/2666253.2666259} (\bibinfo{publisher}{ACM}, \bibinfo{year}{2014}).

\bibitem{Grover2016}
\bibinfo{author}{Grover, S.} \emph{et~al.}
\newblock \bibinfo{title}{Multimodal analytics to study collaborative problem solving in pair programming}.
\newblock In \emph{\bibinfo{booktitle}{Proceedings of the Sixth International Conference on Learning Analytics}}, LAK '16, \bibinfo{pages}{516–517}, \doiprefix\url{10.1145/2883851.2883877} (\bibinfo{publisher}{{ACM} Press}, \bibinfo{year}{2016}).

\bibitem{Cukurova2018}
\bibinfo{author}{Cukurova, M.}, \bibinfo{author}{Luckin, R.}, \bibinfo{author}{Mill{\'{a}}n, E.} \& \bibinfo{author}{Mavrikis, M.}
\newblock \bibinfo{journal}{\bibinfo{title}{The {NISPI} framework: Analysing collaborative problem-solving from students{\textquotesingle} physical interactions}}.
\newblock {\emph{\JournalTitle{Computers \& Education}}} \textbf{\bibinfo{volume}{116}}, \bibinfo{pages}{93--109}, \doiprefix\url{10.1016/j.compedu.2017.08.007} (\bibinfo{year}{2018}).

\bibitem{Muller2018}
\bibinfo{author}{M\"{u}ller, P.}, \bibinfo{author}{Huang, M.~X.} \& \bibinfo{author}{Bulling, A.}
\newblock \bibinfo{title}{Detecting low rapport during natural interactions in small groups from non-verbal behaviour}.
\newblock In \emph{\bibinfo{booktitle}{23rd International Conference on Intelligent User Interfaces}}, IUI '18, \bibinfo{pages}{153–164}, \doiprefix\url{10.1145/3172944.3172969} (\bibinfo{publisher}{Association for Computing Machinery}, \bibinfo{address}{New York, NY, USA}, \bibinfo{year}{2018}).

\bibitem{kim2023multi}
\bibinfo{author}{Kim, J.-H.}, \bibinfo{author}{Kim, N.} \& \bibinfo{author}{Won, C.~S.}
\newblock \bibinfo{title}{Multi modal facial expression recognition with transformer-based fusion networks and dynamic sampling} (\bibinfo{year}{2023}).
\newblock \eprint{2303.08419}.

\bibitem{Xu2023}
\bibinfo{author}{Xu, W.}, \bibinfo{author}{Wu, Y.} \& \bibinfo{author}{Ouyang, F.}
\newblock \bibinfo{journal}{\bibinfo{title}{Multimodal learning analytics of collaborative patterns during pair programming in higher education}}.
\newblock {\emph{\JournalTitle{International Journal of Educational Technology in Higher Education}}} \textbf{\bibinfo{volume}{20}}, \doiprefix\url{10.1186/s41239-022-00377-z} (\bibinfo{year}{2023}).

\bibitem{Lubold2014}
\bibinfo{author}{Lubold, N.} \& \bibinfo{author}{Pon-Barry, H.}
\newblock \bibinfo{title}{Acoustic-prosodic entrainment and rapport in collaborative learning dialogues}.
\newblock In \emph{\bibinfo{booktitle}{Proceedings of the 2014 ACM workshop on Multimodal Learning Analytics Workshop and Grand Challenge}}, ICMI ’14, \doiprefix\url{10.1145/2666633.2666635} (\bibinfo{publisher}{ACM}, \bibinfo{year}{2014}).

\bibitem{beck2001}
\bibinfo{author}{Beck, K.} \emph{et~al.}
\newblock \bibinfo{title}{Manifesto for agile software development} (\bibinfo{year}{2001}).

\bibitem{schwaber2002}
\bibinfo{author}{Schwaber, K.} \& \bibinfo{author}{Beedle, M.}
\newblock \emph{\bibinfo{title}{Agile software development with scrum. Series in agile software development}}, vol.~\bibinfo{volume}{1} (\bibinfo{publisher}{Prentice Hall Upper Saddle River}, \bibinfo{year}{2002}).

\bibitem{Souza2020}
\bibinfo{author}{Guizzardi-Silva~Souza, R.}, \bibinfo{author}{Campos, M.} \& \bibinfo{author}{{Araujo Baiao}, F.}
\newblock \emph{\bibinfo{title}{Applying a Collaboration Domain Ontology Pattern Language in Collaborative Editing: (or: Collaborating, as we learned from Ricardo Falbo)}}, \bibinfo{pages}{62--81} (\bibinfo{publisher}{Nemo Uitgeverij}, \bibinfo{year}{2020}).

\bibitem{Castaer2020}
\bibinfo{author}{Castañer, X.} \& \bibinfo{author}{Oliveira, N.}
\newblock \bibinfo{journal}{\bibinfo{title}{Collaboration, coordination, and cooperation among organizations: Establishing the distinctive meanings of these terms through a systematic literature review}}.
\newblock {\emph{\JournalTitle{Journal of Management}}} \textbf{\bibinfo{volume}{46}}, \bibinfo{pages}{965–1001}, \doiprefix\url{10.1177/0149206320901565} (\bibinfo{year}{2020}).

\bibitem{mahnivc2012using}
\bibinfo{author}{Mahni{\v{c}}, V.} \& \bibinfo{author}{Hovelja, T.}
\newblock \bibinfo{journal}{\bibinfo{title}{On using planning poker for estimating user stories}}.
\newblock {\emph{\JournalTitle{Journal of Systems and Software}}} \textbf{\bibinfo{volume}{85}}, \bibinfo{pages}{2086--2095} (\bibinfo{year}{2012}).

\bibitem{Sharp2010}
\bibinfo{author}{Sharp, H.} \& \bibinfo{author}{Robinson, H.}
\newblock \bibinfo{title}{Three `c's of agile practice: Collaboration, co-ordination and communication}.
\newblock In \emph{\bibinfo{booktitle}{Agile Software Development}}, \bibinfo{pages}{61--85}, \doiprefix\url{10.1007/978-3-642-12575-1_4} (\bibinfo{publisher}{Springer Berlin Heidelberg}, \bibinfo{year}{2010}).

\bibitem{Chassidim2018}
\bibinfo{author}{Chassidim, H.}, \bibinfo{author}{Almog, D.} \& \bibinfo{author}{Mark, S.}
\newblock \bibinfo{journal}{\bibinfo{title}{Fostering soft skills in project-oriented learning within an agile atmosphere}}.
\newblock {\emph{\JournalTitle{European Journal of Engineering Education}}} \textbf{\bibinfo{volume}{43}}, \bibinfo{pages}{638--650}, \doiprefix\url{10.1080/03043797.2017.1401595} (\bibinfo{year}{2018}).

\bibitem{vanLaar2017}
\bibinfo{author}{van Laar, E.}, \bibinfo{author}{van Deursen, A.~J.}, \bibinfo{author}{van Dijk, J.~A.} \& \bibinfo{author}{de~Haan, J.}
\newblock \bibinfo{journal}{\bibinfo{title}{The relation between 21st-century skills and digital skills: A systematic literature review}}.
\newblock {\emph{\JournalTitle{Computers in Human Behavior}}} \textbf{\bibinfo{volume}{72}}, \bibinfo{pages}{577–588}, \doiprefix\url{10.1016/j.chb.2017.03.010} (\bibinfo{year}{2017}).

\bibitem{Johnson_Johnson_2014}
\bibinfo{author}{Johnson, D.~W.} \& \bibinfo{author}{Johnson, R.~T.}
\newblock \bibinfo{journal}{\bibinfo{title}{Cooperative learning in 21st century. [aprendizaje cooperativo en el siglo xxi]}}.
\newblock {\emph{\JournalTitle{Anales de Psicolog\'ia / Annals of Psychology}}} \textbf{\bibinfo{volume}{30}}, \bibinfo{pages}{841–851}, \doiprefix\url{10.6018/analesps.30.3.201241} (\bibinfo{year}{2014}).

\bibitem{Fajaryati2020}
\bibinfo{author}{Fajaryati, N.}, \bibinfo{author}{Akhyar, M.} \& \bibinfo{author}{and}.
\newblock \bibinfo{journal}{\bibinfo{title}{The employability skills needed to face the demands of work in the future: Systematic literature reviews}}.
\newblock {\emph{\JournalTitle{Open Engineering}}} \textbf{\bibinfo{volume}{10}}, \bibinfo{pages}{595--603}, \doiprefix\url{10.1515/eng-2020-0072} (\bibinfo{year}{2020}).

\bibitem{Askari2020}
\bibinfo{author}{Askari, G.} \emph{et~al.}
\newblock \bibinfo{journal}{\bibinfo{title}{The impact of teamwork on an organization's performance: A cooperative game's approach}}.
\newblock {\emph{\JournalTitle{Mathematics}}} \textbf{\bibinfo{volume}{8}}, \bibinfo{pages}{1804}, \doiprefix\url{10.3390/math8101804} (\bibinfo{year}{2020}).

\bibitem{ElSofany2014}
\bibinfo{author}{El-Sofany, H.~F.}, \bibinfo{author}{Alwadani, H.~M.} \& \bibinfo{author}{Alwadani, A.}
\newblock \bibinfo{journal}{\bibinfo{title}{Managing virtual team work in {IT} projects: Survey}}.
\newblock {\emph{\JournalTitle{International Journal of Advanced Corporate Learning ({iJAC})}}} \textbf{\bibinfo{volume}{7}}, \bibinfo{pages}{28}, \doiprefix\url{10.3991/ijac.v7i4.4018} (\bibinfo{year}{2014}).

\bibitem{Dillenbourg2009}
\bibinfo{author}{Dillenbourg, P.}, \bibinfo{author}{J{\"a}rvel{\"a}, S.} \& \bibinfo{author}{Fischer, F.}
\newblock \emph{\bibinfo{title}{The Evolution of Research on Computer-Supported Collaborative Learning}}, \bibinfo{pages}{3--19} (\bibinfo{publisher}{Springer Netherlands}, \bibinfo{address}{Dordrecht}, \bibinfo{year}{2009}).

\bibitem{doi:10.3102/0034654318791584}
\bibinfo{author}{Chen, J.}, \bibinfo{author}{Wang, M.}, \bibinfo{author}{Kirschner, P.~A.} \& \bibinfo{author}{Tsai, C.-C.}
\newblock \bibinfo{journal}{\bibinfo{title}{The role of collaboration, computer use, learning environments, and supporting strategies in cscl: A meta-analysis}}.
\newblock {\emph{\JournalTitle{Review of Educational Research}}} \textbf{\bibinfo{volume}{88}}, \bibinfo{pages}{799--843}, \doiprefix\url{10.3102/0034654318791584} (\bibinfo{year}{2018}).

\bibitem{doi:10.1080/0305764X.2016.1259389}
\bibinfo{author}{Le, H.}, \bibinfo{author}{Janssen, J.} \& \bibinfo{author}{Wubbels, T.}
\newblock \bibinfo{journal}{\bibinfo{title}{Collaborative learning practices: teacher and student perceived obstacles to effective student collaboration}}.
\newblock {\emph{\JournalTitle{Cambridge Journal of Education}}} \textbf{\bibinfo{volume}{48}}, \bibinfo{pages}{103--122}, \doiprefix\url{10.1080/0305764X.2016.1259389} (\bibinfo{year}{2018}).

\bibitem{Eshuis2019}
\bibinfo{author}{Eshuis, E.~H.} \emph{et~al.}
\newblock \bibinfo{journal}{\bibinfo{title}{Improving the quality of vocational students’ collaboration and knowledge acquisition through instruction and joint reflection}}.
\newblock {\emph{\JournalTitle{International Journal of Computer-Supported Collaborative Learning}}} \textbf{\bibinfo{volume}{14}}, \bibinfo{pages}{53–76}, \doiprefix\url{10.1007/s11412-019-09296-0} (\bibinfo{year}{2019}).

\bibitem{Ridao2017}
\bibinfo{author}{Ridao, S.}
\newblock \bibinfo{journal}{\bibinfo{title}{{\guillemotleft}{Es} {un} {lector}, {no} {un} {orador}{\guillemotright}: {Sobre la tricotom\'{i}a comunicaci\'{o}n verbal, paraverbal y no verbal}}}.
\newblock {\emph{\JournalTitle{Revista Digital de Investigaci{\'{o}}n en Docencia Universitaria}}} \textbf{\bibinfo{volume}{11}}, \bibinfo{pages}{177--192}, \doiprefix\url{10.19083/ridu.11.499} (\bibinfo{year}{2017}).

\bibitem{GOSTAND1980}
\bibinfo{author}{GOSTAND, R.}
\newblock \emph{\bibinfo{title}{Verbal and Non-Verbal Communication: Drama as Translation}}, \bibinfo{pages}{1–9} (\bibinfo{publisher}{Elsevier}, \bibinfo{year}{1980}).

\bibitem{Asan2015}
\bibinfo{author}{Asan, O.}, \bibinfo{author}{Young, H.~N.}, \bibinfo{author}{Chewning, B.} \& \bibinfo{author}{Montague, E.}
\newblock \bibinfo{journal}{\bibinfo{title}{How physician electronic health record screen sharing affects patient and doctor non-verbal communication in primary care}}.
\newblock {\emph{\JournalTitle{Patient Education and Counseling}}} \textbf{\bibinfo{volume}{98}}, \bibinfo{pages}{310–316}, \doiprefix\url{10.1016/j.pec.2014.11.024} (\bibinfo{year}{2015}).

\bibitem{Ellgring1981}
\bibinfo{author}{Ellgring, J.~H.}
\newblock \bibinfo{title}{Nonverbal communication - a review of research in germany} (\bibinfo{year}{1981}).

\bibitem{Cockburn2001}
\bibinfo{author}{Cockburn, A.} \& \bibinfo{author}{Highsmith, J.}
\newblock \bibinfo{journal}{\bibinfo{title}{Agile software development, the people factor}}.
\newblock {\emph{\JournalTitle{Computer}}} \textbf{\bibinfo{volume}{34}}, \bibinfo{pages}{131--133}, \doiprefix\url{10.1109/2.963450} (\bibinfo{year}{2001}).

\bibitem{Alsaqqa2020}
\bibinfo{author}{Alsaqqa, S.}, \bibinfo{author}{Sawalha, S.} \& \bibinfo{author}{Abdel-Nabi, H.}
\newblock \bibinfo{journal}{\bibinfo{title}{Agile software development: Methodologies and trends}}.
\newblock {\emph{\JournalTitle{International Journal of Interactive Mobile Technologies ({iJIM})}}} \textbf{\bibinfo{volume}{14}}, \bibinfo{pages}{246}, \doiprefix\url{10.3991/ijim.v14i11.13269} (\bibinfo{year}{2020}).

\bibitem{grenning2002planning}
\bibinfo{author}{Grenning, J.}
\newblock \bibinfo{journal}{\bibinfo{title}{Planning poker or how to avoid analysis paralysis while release planning}}.
\newblock {\emph{\JournalTitle{Hawthorn Woods: Renaissance Software Consulting}}} \textbf{\bibinfo{volume}{3}}, \bibinfo{pages}{22--23} (\bibinfo{year}{2002}).

\bibitem{Wohlin2014}
\bibinfo{author}{Wohlin, C.}
\newblock \bibinfo{title}{Guidelines for snowballing in systematic literature studies and a replication in software engineering}.
\newblock In \emph{\bibinfo{booktitle}{Proceedings of the 18th International Conference on Evaluation and Assessment in Software Engineering}}, EASE '14, \doiprefix\url{10.1145/2601248.2601268} (\bibinfo{publisher}{Association for Computing Machinery}, \bibinfo{address}{New York, NY, USA}, \bibinfo{year}{2014}).

\bibitem{Praharaj_2021}
\bibinfo{author}{Praharaj, S.}, \bibinfo{author}{Scheffel, M.}, \bibinfo{author}{Drachsler, H.} \& \bibinfo{author}{Specht, M.}
\newblock \bibinfo{journal}{\bibinfo{title}{Literature review on co-located collaboration modeling using multimodal learning analytics{\textemdash}can we go the whole nine yards?}}
\newblock {\emph{\JournalTitle{{IEEE} Transactions on Learning Technologies}}} \textbf{\bibinfo{volume}{14}}, \bibinfo{pages}{367--385}, \doiprefix\url{10.1109/tlt.2021.3097766} (\bibinfo{year}{2021}).

\bibitem{Bain2023}
\bibinfo{author}{Bain, M.}, \bibinfo{author}{Huh, J.}, \bibinfo{author}{Han, T.} \& \bibinfo{author}{Zisserman, A.}
\newblock \bibinfo{title}{{WhisperX: Time-Accurate Speech Transcription of Long-Form Audio}}.
\newblock In \emph{\bibinfo{booktitle}{Proc. INTERSPEECH 2023}}, \bibinfo{pages}{4489--4493}, \doiprefix\url{10.21437/Interspeech.2023-78} (\bibinfo{year}{2023}).

\bibitem{10.5555/3618408.3619590}
\bibinfo{author}{Radford, A.} \emph{et~al.}
\newblock \bibinfo{title}{Robust speech recognition via large-scale weak supervision}.
\newblock In \emph{\bibinfo{booktitle}{International Conference on Machine Learning}}, ICML'23 (\bibinfo{publisher}{JMLR.org}, \bibinfo{year}{2023}).

\bibitem{Jocher_Ultralytics_YOLO_2023}
\bibinfo{author}{Jocher, G.}, \bibinfo{author}{Chaurasia, A.} \& \bibinfo{author}{Qiu, J.}
\newblock \bibinfo{title}{{Ultralytics YOLO}}.
\newblock \bibinfo{howpublished}{\url{https://github.com/ultralytics/ultralytics}} (\bibinfo{year}{2023}).
\newblock \bibinfo{note}{License: AGPL-3.0}.

\bibitem{Wojke2018deep}
\bibinfo{author}{Wojke, N.} \& \bibinfo{author}{Bewley, A.}
\newblock \bibinfo{title}{Deep cosine metric learning for person re-identification}.
\newblock In \emph{\bibinfo{booktitle}{2018 IEEE Winter Conference on Applications of Computer Vision (WACV)}}, \bibinfo{pages}{748--756}, \doiprefix\url{10.1109/WACV.2018.00087} (\bibinfo{organization}{IEEE}, \bibinfo{year}{2018}).

\bibitem{mediapipe}
\bibinfo{author}{Lugaresi, C.} \emph{et~al.}
\newblock \bibinfo{title}{Mediapipe: A framework for building perception pipelines}, \doiprefix\url{10.48550/ARXIV.1906.08172} (\bibinfo{year}{2019}).

\bibitem{opencv_solvepnp}
\bibinfo{author}{Bradski, G.}, \bibinfo{author}{Kaehler, A.} \emph{et~al.}
\newblock \bibinfo{title}{Opencv documentation}.
\newblock \bibinfo{howpublished}{\url{https://docs.opencv.org/4.x/d9/d0c/group__calib3d.html}} (\bibinfo{year}{2023}).
\newblock \bibinfo{note}{Accessed: 2024-03-10}.

\bibitem{SHAPIRO1965}
\bibinfo{author}{SHAPIRO, S.~S.} \& \bibinfo{author}{WILK, M.~B.}
\newblock \bibinfo{journal}{\bibinfo{title}{An analysis of variance test for normality (complete samples)}}.
\newblock {\emph{\JournalTitle{Biometrika}}} \textbf{\bibinfo{volume}{52}}, \bibinfo{pages}{591–611}, \doiprefix\url{10.1093/biomet/52.3-4.591} (\bibinfo{year}{1965}).

\bibitem{Levene1960}
\bibinfo{author}{Levene, H.}
\newblock \emph{\bibinfo{title}{Robust Tests for Equality of Variances}}, \bibinfo{pages}{278--292} (\bibinfo{publisher}{Stanford University Press}, \bibinfo{address}{Palo Alto}, \bibinfo{year}{1960}).

\bibitem{kiani2013measuring}
\bibinfo{author}{Kiani, Z. U.~R.}, \bibinfo{author}{Smite, D.} \& \bibinfo{author}{Riaz, A.}
\newblock \bibinfo{title}{Measuring awareness in cross-team collaborations--distance matters}.
\newblock In \emph{\bibinfo{booktitle}{2013 IEEE 8th international conference on global software engineering}}, \bibinfo{pages}{71--79} (\bibinfo{organization}{IEEE}, \bibinfo{year}{2013}).

\bibitem{Haugen2006}
\bibinfo{author}{Haugen, N.~C.}
\newblock \bibinfo{title}{An empirical study of using planning poker for user story estimation}.
\newblock In \emph{\bibinfo{booktitle}{AGILE 2006 (AGILE'06)}}, \bibinfo{pages}{9--pp}, \doiprefix\url{10.1109/agile.2006.16} (\bibinfo{organization}{IEEE}, \bibinfo{year}{2006}).

\bibitem{Poenel2023}
\bibinfo{author}{Poženel, M.}, \bibinfo{author}{F\"{u}rst, L.}, \bibinfo{author}{Vavpotič, D.} \& \bibinfo{author}{Hovelja, T.}
\newblock \bibinfo{journal}{\bibinfo{title}{Agile effort estimation: Comparing the accuracy and efficiency of planning poker, bucket system, and affinity estimation methods}}.
\newblock {\emph{\JournalTitle{International Journal of Software Engineering and Knowledge Engineering}}} \textbf{\bibinfo{volume}{33}}, \bibinfo{pages}{1923–1950}, \doiprefix\url{10.1142/s021819402350064x} (\bibinfo{year}{2023}).

\bibitem{schweighofer2016effort}
\bibinfo{author}{Schweighofer, T.}, \bibinfo{author}{Kline, A.}, \bibinfo{author}{Pavlic, L.} \& \bibinfo{author}{Hericko, M.}
\newblock \bibinfo{title}{How is effort estimated in agile software development projects?}
\newblock In \emph{\bibinfo{booktitle}{SQAMIA}}, \bibinfo{pages}{73--80} (\bibinfo{year}{2016}).

\bibitem{patterson2019equity}
\bibinfo{author}{Patterson, A.~D.}
\newblock \bibinfo{journal}{\bibinfo{title}{Equity in groupwork: The social process of creating justice in a science classroom}}.
\newblock {\emph{\JournalTitle{Cultural Studies of Science Education}}} \textbf{\bibinfo{volume}{14}}, \bibinfo{pages}{361--381} (\bibinfo{year}{2019}).

\end{thebibliography}

\section*{Acknowledgements}
This research was funded by grant ANID/FONDECYT/REGULAR/1211905. Diego Miranda was supported by the ANID Phd Scholarship ANID 21231737.

\section*{Author contributions statement}

R.N., R.M, conceived the experiment(s),  D.M., J.G. and C.E. conducted the experiment(s), R.N., D.M, J.G, R.M and, C.C. analysed the results. All authors reviewed the manuscript. 

\section*{Competing interests}
The author(s) declare no competing interests.

\end{document}